\documentclass[manuscript,screen,nonacm]{acmart}
\AtBeginDocument{%
  \providecommand\BibTeX{{%
    \normalfont B\kern-0.5em{\scshape i\kern-0.25em b}\kern-0.8em\TeX}}}

\usepackage{tcolorbox}
\usepackage{subfiles}
\usepackage{graphicx}
\graphicspath{{images/}}
\usepackage{multirow}
\usepackage{array}
\usepackage{float}
\usepackage{colortbl}
\usepackage{subcaption}

\usepackage{amsthm,amssymb,amsmath}
\usepackage[ruled,vlined]{algorithm2e}
\usepackage{ulem}
\usepackage{soul}
\usepackage{acronym}

{\addtolength{\itemindent}{2em}}%
{\addtolength{\itemindent}{-2em}}

\newcommand{\edit}[1]{\textcolor{black}{#1}}
\newacro{TE}{thought experiments}
\newacro{CRT}{Critical Race Theory}
\newacro{VSD}{Value Sensitive Design}

\usepackage{xcolor}

\newcommand\red[1]{
  \bgroup
  \hskip0pt\color{red}%
  #1%
  \egroup
}
\authorsaddresses{}

\begin{document}


\title[Thought Experiments in HCI]{Thought Experiments for Conceptual Work: A New Application of a (Very) Old Method}

\author{Leah Hope Ajmani}
\affiliation{%
  \institution{University of Minnesota}
  \city{Minneapolis}
  \state{Minnesota}
  \country{USA}}
\email{ajman004@umn.edu}
\author{Mo Houtti}
\affiliation{%
  \institution{Microsoft}
  \city{Redmond}
  \state{Washington}
  \country{USA}}
\email{}
\author{Eric P. S. Baumer}
\affiliation{%
  \institution{University of Toronto}
  \city{Toronto}
  \state{}
  \country{Canada}}
\email{erb416@lehigh.edu}
\author{Stevie Chancellor}
\affiliation{%
  \institution{University of Minnesota}
  \city{Minneapolis}
  \state{Minnesota}
  \country{USA}}
\email{steviec@umn.edu}

\renewcommand{\shortauthors}{Ajmani et al.}
\renewcommand{\shorttitle}{Thought Experiments in HCI}

\begin{abstract}
In this paper, we propose thought experiments (TEs) as a crucial method for Human-Computer Interaction (HCI) researchers to engage in conceptual work. As an interdisciplinary field, HCI often uses concepts as fundamental building blocks for larger theories. However, the conceptual commitments we make in this process carry normative consequences. TEs are a well-established philosophical method, whereby a hypothetical but tractable scenario logically progresses to a conclusion. We outline TEs as an interrogative method that brings conceptualizations to their normative implications through logical moves. We illustrate the value of thought experiments through two examples: (1) original thought experiments to critique stakeholders in Value-Sensitive Design and (2) Helen Nissenbaum’s use of thought experiments to generate contextual integrity. We discuss how TEs precisely anticipate the potential harms of technologies, allowing HCI to operationalize current calls for increased scrutiny of research ethics and broader impacts.
\end{abstract}




\maketitle

\section{Introduction}\label{sec:introduction}

Human-Computer Interaction (HCI) research relies on {\it concepts}. From the idea of ``the user'' to what constitutes ``interaction,'' concepts serve as guideposts for HCI---anchoring abstract thinking~\cite{simon1973structure}; guiding methods~\cite{Oulasvirta2016-en}; and informing design~\cite{Caragay2024-cn}. Over time, concepts evolve through reasoned debate within the field~\cite{Baumer2017-gc, Suchman1993-bm, Harrison2011-hk}. For example, ``the user'' is a central concept in HCI and initially described a single person directly interacting with a system. Now the concept encompasses generalized and complex notions of how people use technology~\cite{Cooper1995-pa, Ackerman2000-ik, Baumer2017-gc}.

Adopting a particular definition for some concept constitutes making a \textit{conceptual commitment}---decisions about the underlying constructs that influence theoretical and empirical investigations. These commitments can be explicit, such as formulating the user as a single individual interacting with a system~\cite{Card2018-ae}. More often than not, however, conceptual commitments are unstated and underpin ongoing work. For example,~\citet{Chancellor2019-yt} found that the concept of the ``human'' in human-centered machine learning is multi-faceted but often implicitly defined by researchers. Indeed, ample work in HCI challenges the conceptual commitments we make in our practice~\citep[e.g.,][]{Ogbonnaya-Ogburu2020-xw, Baumer2011-or, SengersBoehnerEtAl-2008-DisenchantmentAffect, dourish2004we, Chancellor2019-yt}.

Conceptual commitments carry normative consequences. Normative consequences are moral or values-based outcomes that prescribe what is good, desirable, or what we \textit{ought to do}~\cite{Dewey2022-nz, Bynum2000-ty}. We can describe ought-ness in terms of logical consequence (e.g., because I commit to accessibility, I ``ought'' to use 14pt font), or as moral claims about what is right or wrong (e.g., one ought not to steal)\footnote{Given the subject of this paper, many of our thought experiments elucidate moral claims but thought experiments can interrogate normative ones.}. For HCI, normative consequences are the politics, values, and ethical outcomes of our decision-making~\cite{Shilton2018-mk}. 

If left unchecked, the normative consequences of conceptual commitments can contribute to tangible harm. For example, work in feminism~\cite{Bardzell-2010-FeministHCITaking,Bardzell2011-ln,Bardzell-2018-UtopiasParticipationFeminism} and critical race theory~\cite{Ogbonnaya-Ogburu2020-xw, Abebe2022-hs} for HCI describes how our concepts of ``who matters'' are deeply steeped in societal biases and impact our designs. Such work historically examines the societal and practical consequences of HCI, and then seeks to identify the conceptual commitments that give rise to them. 

Our paper starts from an inverse perspective: {\it what if we could evaluate the costs of conceptual commitments in HCI (perhaps a priori), and logically inspect the practical and societal consequences that the commitment causes?} Concepts are powerful building blocks in HCI. Therefore, a failure to understand how our conceptual commitments bear on research decisions risks allowing unethical practices to go unchecked. By unpacking our concepts---the granular components of HCI---we can understand where our normative senses and subsequent research practices come from. The field needs an interrogative methodology for these conceptual commitments: precise reasoning that takes a conceptual commitment to its logical (and often normative) consequence~\cite{Hintikka2002-ug}.



We contribute \textbf{thought experiments for HCI}: a method for HCI researchers to deductively identify conceptual commitments and their normative consequences. Thought experiments are a subclass of philosophical arguments~\citep{Norton1996-tv} that rely on building hypothetical scenarios to explore a logical problem. They are a tool for deductive reasoning; when using a thought experiment, the author(s) precisely toggle experimental variables and explicitly work through logical steps to arrive at a conclusion. This method advances our understanding of how normative consequences relate to the conceptual commitments underlying our theories. In other words, thought experiments answer the question: \textit{if we adopt this particular conceptual commitment, what normative consequences necessarily follow?}

In this paper, we first outline the necessary components of a thought experiment. Then, we introduce evaluative criteria for assessing the rigor of a thought experiment---tractability, validity, and soundness. We demonstrate three practical applications of thought experiments: (1) interrogating the Value-Sensitive Design formulation of \textit{stakeholder}, (2) testing alternative formulations of \textit{stakeholder}, and (3) generating formulations necessary for \textit{contextual integrity}. We illustrate the application of thought experiments by using them to interrogate stakeholderism in HCI, a concept with implications for human-centeredness, justice, and other normative criteria~\cite{Chancellor2019-yt, Neumann2022-th, Deshpande2022-ip}. The formulation of \textit{stakeholder} in VSD determines who gets considered during the design process and, thus, what constitutes a problem. We also show how thought experiments can be generative for HCI by charting Helen Nissenbaum's~\cite{Nissenbaum2004-og} use of thought experiments to develop contextual integrity, a well-known theory of privacy used in HCI~\cite{Skeba2020-zv, Benthall2017-rw, zimmer2018addressing}. We discuss how researchers and practitioners can use thought experiments to further HCI's goals, such as more precise broader impact statements and ethics critiques in peer review. In short, we offer thought experiments as a reliable, rigorous, and precise method for researchers to interrogate the normative consequences of conceptual commitments.

\section{Related Work}\label{sec:related-work}
\subsection{Background and Overview of Thought Experiments}
Thought experiments have a rich history in diverse disciplines, from Achilles and the Tortoise~\cite{Salmon1999-jg} to Schr\"{o}dinger’s cat~\cite{Schrodinger1935-tt}. Scientists have equally used thought experiments to support propositions or to refute them. Notable examples include Galilieo’s Falling Bodies Problem~\cite{Gendler1998-es}, Einstein's Magnet and Conductor Experiment~\cite{Stachel1987-xw, Norton1991-zt}, and Newton's Bucket~\cite{Brown1995-fw, Newton1850-uu}. In moral philosophy, thought experiments are often used to interrogate moral stances rather than scientific propositions. The Trolley Problem\footnote{https://neal.fun/absurd-trolley-problems/} interrogates utilitarianism~\cite{Thomson1985-zu} (albeit a simplified version of the position), and the Violinist Argument similarly interrogates a utilitarian perspective regarding abortion rights~\cite{Thomson2004-hk}. Thought experiments have been heavily leveraged in computational philosophy and related fields. For example, in 1714, Leibniz proposed the Mill Argument to refute the claim that purely material things like machines can think or perceive~\cite{Leibniz1714-av}. This perspective later evolved into computationalism in artificial intelligence, the idea that mental states and computational states are analogous~\cite{Rey1986-fg}. Over 200 years after Leibniz, John Searle proposed the Chinese Room thought experiment (elaborated in Figure~\ref{fig:chinese-room}) to refute the computationalists of his time~\cite{Searle1980-jx}.
                          
Given this diversity of uses, philosophers have not yet coalesced on a single definition of thought experiments. In the 1990s, the power of thought experiments to create knowledge was widely disputed.~\citet{Norton2004-ah}---a noted empiricist---argued that thought experiments are a specific subclass of argument with unique explanatory power but no epistemic power. Norton writes, ``\textit{if thought experiments produce new information about the world it is because they bring to light the hidden consequences of and relations between facts that we already know''~[p. 3]}.

Similarly, thought experiments explain the moral or physical consequences of our intuitions but are not a method for assessing the truth of those initial intuitions. Alternatively,~\citet{Gendler1998-es} argues that thought experiments can moderate epistemological power by scoping what is logically available. Poetically,~\citet{Brown2011-lw} called thought experiments \textit{``a priori science that happens in the laboratory of the mind'' [p. 1]} claiming that thought experiments hold the same epistemic power as physical ones.

In this paper, we adopt Norton’s Argument View that thought experiments are a particular type of argument designed to explain hidden logical consequences~\cite{Norton1991-zt, Norton1996-tv, Norton2002-av}. It is important to note that this deviates from colloquial definitions of thought experiments as speculative games. We frame thought experiments in alignment with philosophy, their discipline of origin. Thought experiments are a formal interrogative method that reveals the implications of conceptualizations. In this way, thought experiments build on complementary prior work in HCI, as described below.

\subsection{Complementary Work in HCI}
\begin{figure}
    \centering
    \includegraphics[scale=.5]{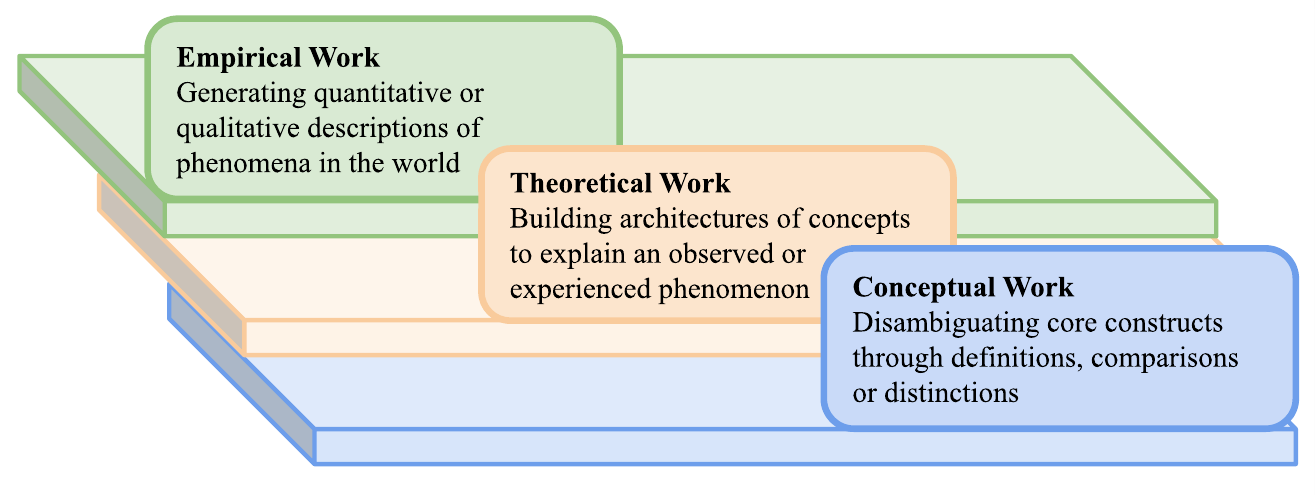}
    \caption{Division of labor between conceptual, theoretical, and empirical work, inspired by~\citet{Oulasvirta2016-en} and~\citet{Kornberger2020-dv}. Note that conceptual work is a base layer that theory and empiricism build on.}
    \label{fig:division}
\end{figure}

\subsubsection{Conceptual Work}
Like many social sciences, HCI welcomes both empirical and theoretical contributions (Figure~\ref{fig:division}). Empirical contributions generate quantitative or qualitative descriptions of phenomena in the world. Meanwhile, theoretical contributions give us architectures of concepts (e.g., frameworks, taxonomies, and models) for understanding our findings. Together, the two inform each other~\cite{Rogers2004-fu}. 

This synergy between theory and empiricism lies on an invisible foundation of conceptual work. As articulated by~\citet{Kirschenmann1982-ia}, concepts are the building blocks of theory. For example, articulating how we conceptualize the human in human-centered machine learning (HCML) is central to building robust theories of data ethics~\cite{Chancellor2019-yt, Ajmani2023-bg}. Therefore, conceptual work must precede theoretical or empirical work. 

A useful concept X has boundary conditions such that we know with some confidence that Y is an instance of X, but Z is not~\cite{Bozeman2007-fc}. Conceptual work is the disambiguation of concepts through definitions, comparisons, or distinctions. \citet{Oulasvirta2016-en} describe conceptual work as the ``problematizing'' stage of HCI research, noting that conceptual problems are second-order: they do not pertain to the world directly nor to our observations or experiences thereof. They argue that papers that propose theories are a form of conceptual contribution. We build on this position of theory generation as a form of conceptual work by shifting focus one step deeper: clarifying and testing consequences of the concepts themselves. HCI has an abundance of theories, many of which are translated from other disciplines, such as feminist theory~\cite{Bardzell-2010-FeministHCITaking}, critical race theory~\cite{Ogbonnaya-Ogburu2020-xw}, and stakeholder theory~\cite{Deshpande2022-ip}. Therefore, the underlying concepts---feminism, power, stakeholder---must be similarly translated. HCI adopts these concepts from their original domains without necessarily interrogating the consequences of conceptual borrowing. We argue that this lack of conceptual methods contributes to a gap in normative reasoning in HCI.

We propose thought experiments as a necessary conceptual method in HCI because they elucidate the edge cases of our conceptual commitments. Here, a weak concept is one where absurd logical conclusions are permissible~\cite{Bozeman2007-fc, Abend2019-lh}; the conceptual commitment is not robust enough to avoid preposterous but logically valid conclusions. In other words, if a conceptualization allows for obviously absurd conclusions,  it indicates there may be a fundamental flaw in the concept itself. In this paper, we specifically show how weak conceptualizations of the stakeholder allow for absurd design choices, such as including clear bad actors (e.g., for-profit prison owners) in the design process. The upshot here is not that researchers are making these choices; it is that they do not have rigorous concepts to justify current practices. The normative implications of our conceptual commitments do not align with our normative intuitions.

\subsubsection{Normative Reasoning}
In philosophy, normative reasoning is the dimension of thinking concerned with ``oughts''~\cite{Sisk2020-xg}. This is distinct from normative reasoning in social and behavioral sciences, which focuses on behaviors that are normalized through societal adoption (e.g., hetero-normativity~\cite{Weathington2023-bx}). Throughout this paper, we use normative in the philosophical sense. 

Normative reasoning in philosophy encompasses ethical and moral reasoning. While the distinction between ethics and morality is a long-standing discourse, 
this paper adopts the stance that ethics defines right and wrong in an environment~\cite{Dewey2022-nz}. Meanwhile, morality defines right and wrong for an individual~\cite{Churchill1982-up}. The intersection of ethics and morals is the sense of right and wrong that informs our actions. This paper focuses on normative statements (i.e., ``oughts'') that lie in that intersection. Specifically, we focus on the implicit oughts that inform design decisions.

HCI has a rich history of ethical and moral reasoning. Prior work has explored ethics---how we ought to do research within our environments---in terms of operationalizing ``isms.'' For example, consequentialism involves thinking of harms and benefits~\cite{Blodgett2022-ky, Shelby2023-ig}, while abolitionism involves interrogating the fundamental purpose behind institutions~\cite{Abebe2022-hs, Liang2021-qb, Kannabiran2023-qd}. This important work lies in the space between theory and empiricism; it involves translating theory into calls to action for the field and its artifacts. Meanwhile, morality in HCI is discussed as researcher positionality. Recent work has called for increased researcher positionality statements~\cite{Chivukula2021-jy, Chancellor2019-cz, Williams2010-bo} to surface how our moral positions impact research decisions. 

In this paper, we take the stance that conceptual positions carry normative consequences. When we make a decision about a concept, we also bring implications for what is morally permissible. Yet, these consequences are difficult to anticipate without the proper methods. We propose thought experiments as an accessible method for HCI practitioners to engage with normative reasoning that often goes unchecked in HCI.

\subsubsection{Distinction from Speculative Methods}
HCI researchers and practitioners may be inclined to draw parallels between thought experiments and speculative methods, given that the latter also involves reasoning about hypothetical scenarios. Indeed, some prior work has claimed that speculative methods such as design fiction can be conceived of as thought experiments~\citep[e.g.,][]{Baumer2020-ol, Barendregt2021-or, Blythe2018-vd, Bleecker2009-io}. However, we caution against such comparisons. Thought experiments and speculative methods leverage hypotheticals in entirely different ways for entirely different purposes. 

Speculative methods like futuring~\cite{Harrington2022-tj}, design fiction~\cite{Coulton2017-oc}, and design probing~\cite{Noortman2019-vs} invite us to envision alternative realities and futures, deriving inspiration from storytelling and fiction~\cite{Russell2018-qi}. They often focus on being provocative~\cite{Bray2022-zj}, critical~\cite{Pierce2015-ww}, or radically imaginative~\cite{Coulton2017-oc}. By contrast, thought experiments use hypotheticals to create logical guardrails for testing a proposition deductively~\cite{Norton1996-tv, Gendler1998-es, Brown1995-fw}. This logical interrogation, originating from a separate tradition rooted in philosophy, allows us to precisely evaluate propositions, including propositions about concepts. 

To illustrate this difference, we can look to the futures cone: a conical projection of preferable, probable, plausible, and possible futures~\cite{Hancock1994-tx, Gall2022-ww}. In speculative methods, participants are tasked with traversing the time axis and building hypothetical worlds~\cite{Dunne2013-un}. In thought experiments, there is no concept or categorization of futures. The author picks an imagined scenario that captures the relevant dimensions, regardless of its likelihood as a possible future. Once the author posits a hypothetical, there is no world-building, only deducing. The crucial contribution of a thought experiment is the logical conclusion one makes, not the hypothetical world one built. While superficial similarities may exist, looking beneath the surface reveals that speculative methods and thought experiments are highly dissimilar in their purposes and intellectual origins.

\subsection{Our cases}
\subsubsection{Stakeholderism in HCI}
In 1984, Freeman introduced the stakeholder concept in Organizational Sciences as a new way of framing ideal corporate strategy for firms~\cite{freeman1984strategic}. Stakeholder theory asserts that the firm is a network of value-creation obligations for those who can affect or be affected by the realization of an organization’s purpose (the wide definition) or those without whose support the organization would not exist (the narrow definition)~\cite{Dmytriyev2021-lj}. In the past 40 years, stakeholder theory has become widely accepted and iterated on in organizational research. For example, stakeholder theory has grounded corporate social responsibility~\cite{Dmytriyev2021-lj}, business ethics~\cite{Gibson2000-zd}, and non-financial value creation~\cite{Hatherly2020-vg}. Moreover, stakeholder theory transformed how managers and corporations are evaluated. ~\citet{Clarkson1995-we} notes that, under stakeholder theory, firms should be evaluated by ``\textit{the ability of its managers to create sufficient wealth, value, or satisfaction for those who belong to each stakeholder group.}"

In HCI and CSCW, stakeholder theory has been heavily adapted for systems and applications. In this sense, the firm is a system or application, and a stakeholder is someone who should be considered in the design process. Similar to organizational research, the term stakeholder carries moral weight; when we fail to include stakeholders in the design process, we are skirting a moral obligation to do right by them~\cite{Zhu2018-ov, Neumann2022-th, Chancellor2023-mw}. Researchers agree that it is imperative to better understand the term stakeholder~\cite{Zhang2023-ta, Zhang2023-hf, Deshpande2022-ip}. Previous work has shown that excluding specific stakeholder considerations from the design process can harm vulnerable populations. For example,~\citet{Stapleton2022-bu} found that including parents in predictive child-welfare algorithm design fosters a richer understanding of the algorithm's potential shortfalls.~\citet{Chancellor2019-yt} found that predictive mental health researchers have not coalesced on a conceptualization of stakeholders; data subjects are conceptualized as a myriad of different roles, from patients to social media users. This discord can cause dramatic differences in data handling procedures and limit human agency in sociotechnical systems~\cite{Bardzell2015-zf}.  

This paper uses thought experiments to interrogate a widely accepted conceptualization of stakeholders in HCI through Value-Sensitive Design (VSD). VSD states that understanding relevant community stakeholders’ motivations, values, and goals is an essential first step of the design process~\cite{Friedman1996-kx}. Subsequently, VSD conceptualizes stakeholders as anyone who either (1) interacts with a system or output or (2) is affected by a system or output. In VSD, (1) is a direct stakeholder, while (2) is an indirect stakeholder. However, previous work has noted limitations of this conceptualization. ~\citet{Baumer2017-gc} note that stakeholder-system relationships do not follow straightforward patterns.~\citet{Borning2012-ki} note that researcher positionality is often understated in VSD research papers. Furthermore, identifying ``key'' stakeholders relies on researcher rationale that is often overlooked~\cite{Yoo2017-hx}. We build on these critiques by proposing various expansions of VSD and alternative conceptualizations. We use thought experiments to demonstrate that these alternatives better achieve the moral imperatives of stakeholderism in HCI.

\subsubsection{Privacy and Contextual Integrity in HCI}
In 2004, Helen Nissenbaum presented Contextual Integrity (CI) as a tool for understanding privacy~\cite{Nissenbaum2004-og}. CI asserts that adequate privacy is provided when an information flow conforms to both implicit and explicit norms surrounding information sharing. CI has since been formalized into five parameters: (1) the information subject; (2) the sender; (3) the recipient; (4) the information type; and (5) the transmission principles that represent the norms of information flow~\cite{Barth2006-bu}. A contextual integrity analysis requires a practitioner to define each of these parameters in relation to the context they are assessing. Notably, CI is a large departure from previous privacy theories, such as privacy as control~\cite{Posner1978-mz}, secrecy~\cite{Warren1977-zc}, or information security~\cite{Thompson2001-tf}. 

While not originally published in HCI venues, Nissenbaum and her peers have inspired radical change to both HCI research and privacy policy surrounding data sharing. For example,~\citet{zimmer2018addressing} leveraged contextual integrity and its metaphors to guide ethical decision-making for research using big data. Policy-oriented institutions, such as the Federal Trade Commisssion, have cited Nissenbaum's work as grounds for new laws around data privacy~\cite{Madrigal2012-is}. Moreover, CI has sparked numerous open questions.~\citet{Chancellor2019-cz} ask if contextual integrity can be applied to online mental health communities that may not want to be discovered.~\citet{Skeba2020-zv} ask if contextual integrity is still relevant to inferential data.

We elaborate on Nissenbaum's methodology for generating contextual integrity. We highlight the thought experiments that Nissenbaum and her peers used throughout their iterations of CI. By formalizing Nissenbaum's implicit use of thought experiments into our explicit structure, we show that thought experiments can be conceptually generative---giving us formulations that serve as the building blocks for larger theories.

\section{How to Conduct a Thought Experiment}\label{sec:how-to}
In philosophy, there is ample discourse around the different taxonomies, implications, and specifics of thought experiments \citep[e.g.][]{Gendler1998-es, Brown1995-fw, Norton1996-tv, Sorensen1992-vq}. These discussions rely on a shared understanding of thought experiments to guide a proper philosophical analysis~\cite{Haggqvist2009-bw}. In this section, we explain common features of building and evaluating a thought experiment. We propose a definition of thought experiments that is both true to the centuries of philosophy research and applicable to HCI: thought experiments are a type of argument that (1) posit a hypothetical and (2) invoke experimental conditions as a means to interrogate the normative implications of conceptual commitments~\cite{Norton1996-tv}. Thought experiments require both the author and the reader to adopt a philosophical stance---a space between the~\edit{intuitive} and empirical where experimental hypotheticals are crucial reasoning devices~\cite{Kornberger2020-dv}.
\subsection{Components of Thought Experiments}
As~\citet{Brown1995-fw} articulates, engaging with a thought experiment involves these general steps: 
\begin{enumerate}
    \item Postulating a proposition or counterfactual
    \item Visualizing some situation that we have set up in the imagination
    \item Carrying out an operation to see what happens
    \item Drawing a conclusion
\end{enumerate}
Each of these steps relies on a rhetorical component:
\begin{enumerate}
    \item \textbf{Proposition.} A claim or stance the author puts forth to motivate the thought experiment~\cite{Norton2004-ah}.
    \item \textbf{Analogy.} The imagined hypothetical that acts as a physical setting for the subsequent ``experiment."~\cite{Mach1893-sz}
    \item \textbf{Experimental Variable.} An experimental dimension the author varies to further their argument~\cite{Gendler1998-es}.
    \item \textbf{Conclusion.} A claim about the initial proposition that logically follows from the thought experiment.
\end{enumerate}

A complete thought experiment adopts a stance and reasons about the implications of changing specific dimensions within a hypothetical world. Thus, thought experiments take a proposition to its logical conclusion. \edit{In this paper, we present thought experiments in both prose and formal logic for comprehensiveness. However, expertise in formal logic or the technical logistics of philosophical argumentation is not necessary to produce a high-quality thought experiment. Many famous scientists have presented seminal thought experiments in prose only (e.g., John Searle's Chinese Room Thought Experiment~\cite{Searle1980-jx}).}

\begin{figure}
    \centering
    \includegraphics[width=\textwidth]{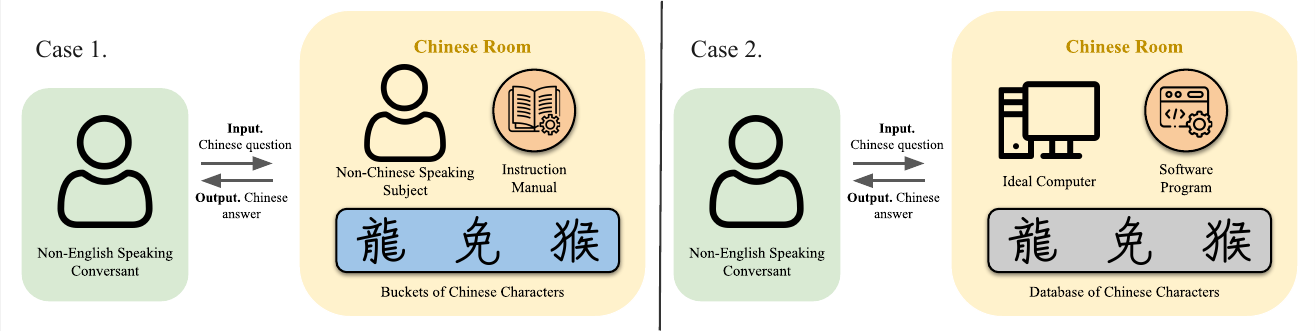}
    \caption{\edit{Visual of Searle's Chinese Room thought experiment. We present two cases: (1) the basic setup of a human in the Chinese room and (2) the experimental case of the ideal computational machine.} While both cases would simulate conversation, Searle concludes that the human could reach a mental state of understanding that a machine could never achieve.}
    \Description{Fig 1. Visual of Searle's Chinese Room thought experiment depicting two cases. Case 1 - a non-English peaking conversing passing questions into the Chinese room. In this case, the Chinese room contains a human subject, an instruction manual, and buckets of Chinese characters. Case 2- the Chinese room now contains a computer with a program for answering Chinese questions.}
    \label{fig:chinese-room}
\end{figure}

\subsubsection{Example. }In 1980,~\citet{Searle1980-jx} proposed the Chinese Room thought experiment as a refutation of his computationalist contemporaries. At the time, AI hype was causing optimism and panic. By the late 1970s, some AI researchers claimed that computers already understood at least some natural language~\cite{Simon1977-gv}. To this day, Searle's Chinese Room is heralded as a seminal use of thought experiments in computational fields~\cite{Preston2002-ix}. In this section, we walk through Searle's Chinese Room~\cite{Searle1980-jx} to exemplify how to formally conduct this common thought experiment. 

\textbf{Proposition. } To start, we seek to refute a computational conceptualization of the mind: the human mind is a program being run on the brain in the same way that software is a program that gets run on a \edit{computational processing unit}. In popular discourse, this implies the existence of \textit{Strong AI}, a computer that, given unlimited resources, could replicate every task of the human brain. More formally,
\begin{proof}\leavevmode
    \begin{enumerate}
    \item The mind is a program run on an extremely complex machine (i.e, the brain) \hfill(Definition of Mind)
    \item Given the right resources, a computer can perform every task the human mind can\hfill (Implication)
    \end{enumerate}
\end{proof}

\textbf{Analogy. }Imagine we, a native English speaker who knows no Chinese, are asked to converse with a native Chinese speaker who knows no English. However, we are in a room (henceforth, The Chinese Room) that has buckets of Chinese symbols and a book of instructions for how to manipulate the symbols (see Figure~\ref{fig:chinese-room}).

Suppose our conversant is outside this room, passing us questions written in Chinese. By properly following the instruction manual, we can manipulate this input and produce a sequence of Chinese characters that correctly answers their question. We can infinitely continue this process of engaging in ``conversation.'' However, Searle notes that this conversation does not demand that we (the human) understand Chinese~\cite{Searle1980-jx}. Under this computational representation of conversation, a human can exist in the Chinese Room and never learn a single word of Chinese.

\textbf{Experimental Variable. }Suppose instead of placing ourselves in the Chinese Room, we place \edit{an unlimited processor} in the room. Note the parallels in our initial analogy to a computer-based situation. Instead of a room with buckets of symbols, there is a database of Chinese characters; instead of an instruction manual, the \edit{processor} is given a program for transforming Chinese. \edit{Essentially, the Chinese Room is now a computer.} We can then picture the conversation task as a Chinese speaker passing in a question input and receiving an answer output.

\textbf{Conclusion. }In our original analogy, we do not arrive at a greater understanding of Chinese. In fact, we could answer an infinite number of questions from the conversant and not understand a single word of Chinese. However, we would be simulating conversation by accurately answering Chinese questions.

This conclusion holds in our computer-based case as well. If we do not understand Chinese based on implementing the appropriate program for speaking Chinese, neither does any other digital computer solely on that basis. Note that the computer, by our conceptualization of the mind, does not have anything the human does not have.

At this point, we have reached a logical contradiction: there is a state of the mind (i.e., understanding Chinese) that a functionally described machine, either human or ~\edit{computational}, could never have. This contradicts our original conceptualization of the human mind as a sophisticated program run on the brain. By definition, programs are input manipulation tasks. However, the Chinese Room experiment demonstrates that one could manipulate symbols for eternity and never reach \textit{understanding}. Therefore, there exists a human mental state that is impossible to describe under a computational conceptualization of the mind. Subsequently, the computational conceptualization of the mind does not hold. In formal logic, the Chinese Room argues:
\begin{proof}\leavevmode
    \begin{enumerate}
    \item $mind = program(Brain)$\hfill(Definition of Mind)
    \item $\forall \text{ tasks } t,\ mind \text{ can complete }t \rightarrow computer \text{ can complete }t$ \hfill(Implication)
    \item $Let\ t = \text{understanding Chinese}$
    \item $mind \text{ can complete }t$ \hfill
    \item $computer \text{ cannot complete }t$ \hfill
    \item $\exists \text{ task } t \text{ such that } mind \text{ can complete }t \land computer \text{ cannot complete }t$ \hfill(Contradiction of 2)
    \item $mind \neq program(Brain)$
    \end{enumerate}
\end{proof}

\subsubsection{Scoping Thought Experiments.} As with empirical experiments, thought experiments have guardrails that scope the realm of imagined possibility. In speculative design methods, these guardrails have been described as the ``futures cone'': a conical projection of preferable, probable, plausible, and possible outcomes~\cite{Gall2022-ww}. Similarly, the author must define and defend a reasonable realm of possibility in thought experiments. For physics thought experiments, such as Galileo’s Falling Bodies problem, the realm of possibility is all previously proven physics principles. In Searle’s Chinese Room, one can only communicate by passing notes off Chinese characters. These guardrails help make ``instinctive knowledge"~\cite{Mach1893-sz} (i.e., intuitions) a permissible explanatory device. Because the author has built a world with specific dimensions, they have scoped what is logically available to the reader. This technique allows readers to interrogate how their intuitions interact with this hypothetical scenario. The realm of possibility in thought experiments is crucial because it allows readers to rely on their intuitions while also interrogating them.

\subsection{Evaluating a Thought Experiment}

Like most logical arguments, a high-quality thought experiment must engage in high-quality reasoning. Building on formative work about thought experiments~\cite{Mach1893-sz, Norton2002-av, Brown1995-fw}, we describe three evaluative criteria for thought experiments: tractability, validity, and soundness. Tractability evaluates whether the hypothetical is appropriate for the argument being made. Meanwhile, validity and soundness evaluate the logical moves themselves.
\subsubsection{Evaluating the Analogy.} Transforming a problem into a new solution space necessitates creating a tractable version of the initial problem statement. However, this transformation creates the potential for faulty representations that strong-arm or strawman a problem statement into an inappropriate solution space. \citet{Baumer2011-or} describe this in HCI as \textit{``solving a computationally tractable solution to the problem rather than the problem itself.”} Thought experiments suffer from a similar problem: authors can build analogies that do not parallel the real dilemma being interrogated. We clarify and explain how to evaluate thought experiments by contrasting two principles---tractability and realizability~\cite{Mach1893-sz}.

\textbf{Tractability describes whether the imagined scenario properly interrogates the proposition without begging for a trivial logical argument.} Tractability is crucial for useful thought experiments with tangible implications. However, it is important to distinguish between \textit{tractability} versus \textit{realizability} when evaluating thought experiments. While the imagined scenario should be relevant to the real world, it does not have to perfectly mirror a real-world experimental setup. Examining tractability in thought experiments is a common mechanism for critically evaluating them. For example, critics of the Chinese Room have noted that modern computer programs can interact with the real world, which could potentially lead to understanding (see ``The Robot Reply"~\cite{Rodriguez2012-mw, Dennett1980-qg, Rey1986-fg}). Therefore, a computer in a windowless room is not tractable to modern programs, so the thought experiment's conclusion is irrelevant to the original proposition. Searle has responded to these critics by expanding his thought experiment to include real-world interaction and still reaching similar conclusions~\cite{Bishop2001-mz}. Note that avid criticism in philosophy is often the sign of a provocative argument rather than a highly flawed one, so debates like these do not indicate that Searle's argument should be rejected.  

Realizability---the ability for a thought experiment to be replicated in a world free of physical, ethical, time, or financial limitations---is \textbf{not} necessary for a useful thought experiment.~\citet{Mach1893-sz} argues that physical unrealizability need not be a defining condition of thought experiments. Unrealizable amounts of money or fictitious characters are not reasons why thought experiments would be rejected; preposterous exaggerations may capture a concept well. Mach describes this as the core of taking a philosophical stance: sitting in the space between logic and instinctive knowledge. Therefore, thought experiments should not be evaluated on how well they would replicate in a condition-less world. Rather, they should be evaluated on how well they parallel certain ideas and give them understandable constructs. For example, Searle represents the idea of computational omniscience with a program that can fluently converse in Chinese.

For an exaggerated example of a thought experiment analogy that makes a conclusion trivially true, we could change the task in Searle's Chinese Room from conversing in Chinese to a non-language-based task. Suppose instead we change the input to several digits $n$ and the output to the $n$-th digit of Pi. The human in the room is given a book with all of the digits of Pi (note that while physically impossible because Pi is infinite, this is logically plausible). In essence, instead of Searle's Chinese Room we have Searle's Pi Room. In this scenario, there is no deeper level of \textit{understanding} that is available to the human. It is not the case that a human can reach a mental state in this task that a computer cannot. Therefore, we would no longer reach a logical contradiction from our initial conceptualization of Strong AI.

Searle's Pi Room is a poorly designed thought experiment because the task of finding a digit of Pi is an inherently computational task. Therefore, the conclusion that a computer and human would complete it analogously is trivial. There is no notion of ``understanding Pi" in the same way that there is a notion of ``understanding Chinese." Therefore, this analogy is untractable; humans obviously complete a myriad of tasks that are non-computational in nature, such as learning a new language. Searle's Pi Room does not parallel the real-world dilemma of computationalism in relevant ways, which, in turn, causes it to rely on trivial logic.

\subsubsection{Evaluating The Argument.} Recall that we adopt Norton’s view that thought experiments are a specific type of argument. Therefore, they are subject to the same type of evaluation as other reasoning methods, like proofs and deductive logic. Particularly in philosophy, evaluating logical arguments has a rich history with two evaluation criteria~\cite{Caret2015-rh, Lloyd2012-fu, Hintikka2002-ug}: (1) validity and (2) soundness. 

{\bf Validity describes the quality of each step in the argument; an argument is \textit{valid} if the truth of the premises logically guarantees the truth of the conclusion.} For example, the Chinese room experiment takes a common logical form, \textit{reductio ad absurdum}. Searle posits: if Strong AI is true, then any human mind state can be achieved by a computational program. Through a thought experiment, he reaches the logical contradiction that human understanding cannot be simulated by a computational system, whether that system is human or mechanic. In other words, Searle is able to refute the conceptualization of Strong AI by showing that accepting it would lead us to an absurd conclusion. Thought experiments, given their hypothetical nature, provide the tools needed to set up an illustrative case that might not occur in the real world. These cases are often edge cases that are difficult to reason about in informal ways. By taking logical steps and ending up at an absurd conclusion, Searle demonstrates a flaw in the initial conceptualization.

An invalid thought experiment means that the author has used an unjustified logical move to progress the thought experiment. For an exaggerated example, Searle's logic would be invalid if he had used the claim (1) ``I could run a program for Chinese without thereby coming to understand Chinese" to deduce (2) ``therefore, no computer can converse in Chinese." Jumping from claim (1) to claim (2) would require that Searle establishes an equivalence between ``understanding" and ``conversing." However, this claim would undermine the initial setup that one can converse in Chinese if given the right tools. Therefore, this logical jump would make the argument invalid.

{\bf Soundness describes whether the initial premises are, in fact, true.} Critiquing the soundness of a thought experiment necessitates critiquing whether the initial premises accurately represent the conceptualization being interrogated. Thought experiments rely on formalizing arguments in the aether into logical representations of them.

We identify thought experiments where the initial premise is a conceptual commitment made by translating theory into HCI or a commitment made through a design. In this respect, soundness in thought experiments describes whether the initial premise is a fair, logical representation of the concept or situation. For example, critics of Searle argue that he takes a reductivist conceptualization of computationalism (see ``The Intuition Reply"~\cite{Block1995-gl, Sprevak2007-rs}). These critiques are valuable because they allow us to consider the parallels between the thought experiment and popular discourse about concepts. Searle responded to these critics by noting that many prominent figures, like Herbert Simon, already claim that we have machines that can think~\cite{Searle1984-jf}. This response highlights that Searle's thought experiment accurately represents at least one flavor of the computationalist position he is interrogating. Therefore, Searle's Chinese Room is argued to be fairly sound. This example shows that soundness critiques can illuminate useful discussions about the status quo and the pragmatic implications of conceptualizations in popular discourse.

\subsubsection{\edit{Normative Consequences of Thought Experiments}}
It is important to recognize that using thought experiments carries its own normative consequences. As with any research method, engaging with thought experiments presumes a particular epistemic stance. We adopt~\citet{Norton2002-av}'s view that thought experiments do not transcend empiricism. For example, replacing an interview study with a thought experiment that speculates, \textit{``What would these stakeholders think?''} is epistemically invalid. Researchers cannot gain knowledge about people's perceptions through thought experiments. Thought experiments are designed to answer conceptual research questions that inform theoretical and empirical ones.

Moreover, thought experiments do not create normative guarantees. An argument presented as a thought experiment may be logically valid and sound but have a problematic conclusion. For example, Iris Young notes that the outcome of Rawls's famous Veil of Ignorance thought experiment is an unacceptable formulation of justice, as it discounts systemic oppression~\cite{Young1981-np}. Put concisely, thought experiments do not necessarily guarantee ethical conclusions. Instead, thought experiments make the implicit both explicit and precise to ground further deliberation.

\section{Interrogating the VSD Formulation of Stakeholders}\label{sec:vsd-stakeholder}

\begin{figure}
    \centering
    \includegraphics[width=\textwidth]{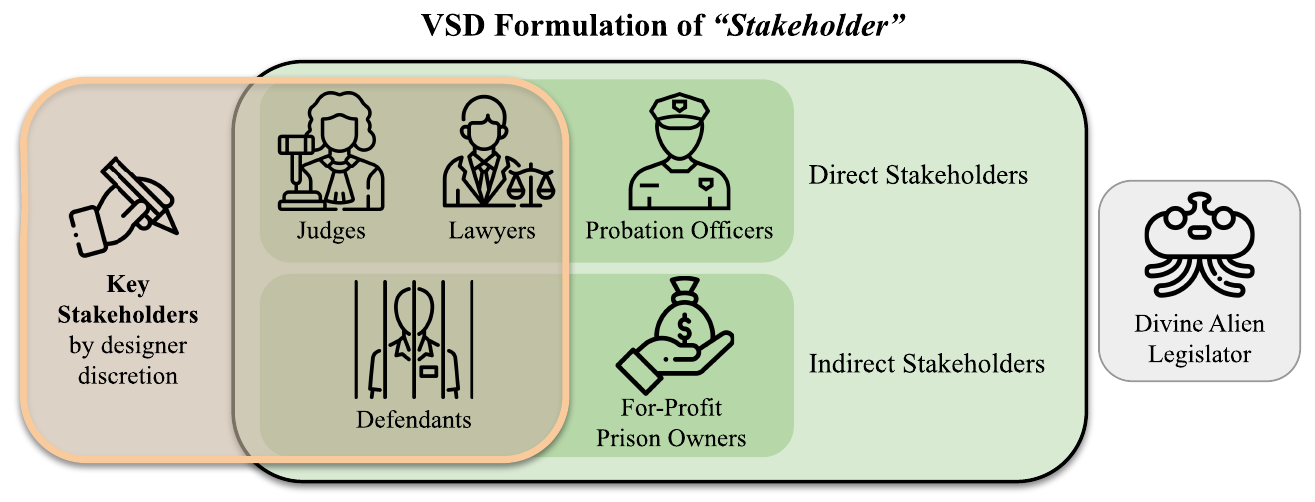}
    \caption{Visual of VSD's formulation of stakeholders. Note that VSD relies on three different sets of stakeholders: (1) direct stakeholders who interact with the system, (2) indirect stakeholders who are affected by the system, and (3) key stakeholders who are identified by the practitioner as stakeholders whose perspectives are substantially important~\cite{Friedman1996-kx}.}
    \Description{Fig 2. Visual of VSD's formulation of stakeholders. Image depicts judges, lawyers, and probation officers as direct stakeholders. Defendants and for-profit prison owners are separated as indirect stakeholders. The divine alien legislator is depicted outside the bucket of stakeholders. Judges, lawyers, and defendants are highlighted with an orange box as key stakeholders by designer discretion.}
    \label{fig:vsd-formulation}
\end{figure}

In this section, we present several thought experiments to interrogate the concept of stakeholders in VSD. First proposed by~\citet{Friedman1996-kx}, Value-Sensitive Design (VSD) is a methodology that asserts system designers should understand values early in the design process to build~\edit{usable} and ethical systems. Recall from our Related Work (Section~\ref{sec:related-work}) that, under VSD, stakeholders are either \textit{direct} or \textit{indirect} based on their relationship to a system or output.

As VSD researchers have noted, this is an extremely general conceptualization of stakeholders. In theory, every human on the planet could be considered an indirect stakeholder~\cite{Davis2015-fj, Borning2012-ki}. Therefore, as VSD methods have become more popular, researchers have augmented VSD's initial stakeholder definition, specifying that key stakeholders are people ``\textit{who are or will be significantly implicated by the technology.}''~\cite{Friedman2013-ge}

Since stakeholder prioritization is left to the researcher's discretion in VSD, it is important to have robust methods that \edit{unpack researcher intuitions (see Figure~\ref{fig:vsd-formulation}). Unchecked} intuitions about who gets to be included in the design process have caused severe harm, often to communities that are already marginalized~\cite{Chancellor2019-yt, DeVito2021-yv, Stapleton2022-bu,Cheng2022-dv}. 

In this section, we ask: 
\textit{What are the normative consequences of committing to the stakeholderism established in VSD?}

We use thought experiments to demonstrate that the stakeholder concept in VSD creates room for problematic normative judgments that do not align with previous work~\cite{Hallett2006-qu}.

\subsection{Experiment Setup} We begin by setting up a tractable situation where stakeholderness has significant normative implications.

\textbf{Proposition. }
Broken down into premises, VSD states:
\begin{proof}\leavevmode
    \begin{enumerate}
    \item $P \in \{stakeholders\} \iff (P \text{ interacts with } T \lor P \text{ is affected by } T)$ \hfill(VSD-definition of stakeholder)
    \item $P \text{'s perspective should be valued } \iff P \in \{stakeholders\}$ \hfill (VSD-implication of stakeholder inclusion)
    \end{enumerate}
\end{proof}

\textbf{Analogy. }Suppose we have a technical system $T$, designed to predict a defendant's recidivism probability in court. $T$'s primary use is for judges to determine proper sentencing for a convicted defendant. We use the social role of person $P$ as our experimental variable and represent the set of all stakeholders of $T$ as $\{stakeholders\}$.~\edit{We present the first few thought experiments in both prose and formal logic. Recall that formal logic is not necessary for thought experiments. We demonstrate a prose-only thought experiment in Section~\ref{sec:dal-te}.}

\subsection{Basic Example} To start, we provide a straightforward thought experiment that shows VSD's conceptualization of stakeholder inclusion justifies consulting the primary user (i.e., judges) in the design process.

\textbf{Experimental Variable. }Let $P$ be a judge who will use $T$. Given $P$'s job as a judge, we know that $P$ will interact with $T.$ Therefore, $P$ meets VSD's criteria for stakeholder inclusion. 

\textbf{Conclusion. }Because $P$ is a stakeholder of $T$, $P$'s perspective should be valued in designing $T.$ Written formally,
\begin{proof}\leavevmode
    \begin{enumerate}
    \item $P := \text{judge}$\hfill(Given)
    \item $P \in \{stakeholders\} \iff (P \text{ interacts with } T \lor P \text{ is affected by } T)$ \hfill(VSD-definition of stakeholder)
    \item $P \text{'s perspective should be valued } \iff P \in \{stakeholders\}$ \hfill (VSD-implication of stakeholder inclusion)
    \item $P \text{ interacts with } T$ \hfill(From 1)
    \item $P \in \{stakeholders\}$ \hfill(From 2, 4)
    \item $P$\text{'s perspective should be valued}\hfill(From 3, 5)
    \end{enumerate}
\end{proof}

This thought experiment demonstrates that the VSD conceptualization of a stakeholder justifies consulting judges during the design process. However, we can start to reveal more convoluted normative consequences by altering $P$'s role.

\subsection{Conceptualization is Over-Inclusive} 
Next, we show that VSD deviates from normative prescriptions (i.e., calls for what \textit{ought} to be) by being over-inclusive. Said another way, we use a thought experiment to show that people who, based on prior work~\cite{Angwin2016-jy, Cohen2015-ev, Hallett2006-qu}, should be intentionally excluded in the design process are justified stakeholders under VSD. This over-inclusiveness implies that the conceptualization does not align with what \textit{ought} to be the case. In other words, VSD allows for absurd stakeholder choices. The following thought experiment demonstrates the normative consequence of said commitment:

Let $P$ be a for-profit prison owner. Given $P$'s social role, $P$ will be heavily affected by $T.$ For example, if $T$ over-predicts recidivism, more offenders could be sent to prison, and $P$'s capital would increase. Therefore, $P$ meets VSD's criteria for stakeholder inclusion. Because $P$ is a stakeholder of $T$, $P$'s perspective should be valued in designing $T.$ 
\begin{proof}\leavevmode
    \begin{enumerate}
    \item $P := \text{for-profit prison owner}$\hfill(Given)
    \item $P \in \{stakeholders\} \iff (P \text{ interacts with } T \lor P \text{ is affected by } T)$ \hfill(VSD-definition of stakeholder)
    \item $P \text{'s perspective should be valued } \iff P \in \{stakeholders\}$ \hfill (VSD-implication of stakeholder inclusion)
    \item $P \text{ is affected by } T$ \hfill(From 1)
    \item $P \in \{stakeholders\}$ \hfill(From 2, 4)
    \item $P$\text{'s perspective should be valued}\hfill(From 3, 5)
    \end{enumerate}
\end{proof}
This thought experiment demonstrates that a for-profit prison owner is a stakeholder. Researchers could justifiably incorporate $P$'s perspective into the design process. Moreover, researchers have no normative justification to exclude $P$ from being a stakeholder. However, valuing $P$'s perspective in the design process would be inconceivable. Most would balk at including a for-profit prison owner in a values-sensitive exchange, as prioritizing capital interests has historically caused significant harm and exploitation in prison systems~\cite{Hallett2006-qu, Cohen2015-ev}. Note that our current conceptualization of stakeholders has no inclusion or exclusion criteria concerning purely capital interests. Essentially, VSD is over-inclusive; monied claims can justify stakeholdership. However, nothing in the conceptualization makes this hierarchy morally reprehensible. 

In this thought experiment, we demonstrate that the VSD concept of a stakeholder is not robust enough to guide the decisions that researchers need to make. In effect, there is a misalignment between the concept's normative implications and the field's normative intuitions. The problem here is not that a designer will now think it is moral to include a for-profit prison owner as a stakeholder. Rather, the term ``stakeholder'' does not properly capture our intuitions on what ought to be. Therefore, the outcome of $P$ as a for-profit prison owner reveals that we need to bound the conceptualization of stakeholders to separate moral obligations from financial ones.

\subsection{Conceptualization is Under-Inclusive}~\label{sec:dal-te}
Next, we show how VSD simultaneously \textit{narrows} the definition of stakeholder. Similar to the above thought experiment, we show that VSD’s conceptualization of stakeholders is bounded implicitly rather than explicitly. By committing to VSD’s definition, we draw a boundary that excludes specific stakeholders from consideration.

Let $P$ be a legitimately elected legislator who votes on whether $T$ gets used in the judicial system. Suppose, because of their role as a legislator rather than a judge, $P$ will never interact with $T.$ Therefore, $P$ is a stakeholder of $T$ if and only if $P$ is \textit{affected by} $T.$ Yet legislators, we posit, are not likely to be meaningfully affected by the recidivism prediction system. Therefore, by VSD, $P$ is \textit{not} a stakeholder of $T$. Accepting VSD’s conceptualization of stakeholders precludes us from needing or seeking to consult legislators when designing systems—even with major public policy implications. 

We pause to address a potential question: \textbf{are legislators not affected by the system?} For one, legislators are answerable to their constituents---and unhappy constituents are not likely to re-elect them for a second term. Therefore, if a system affects the legislator’s constituents, it involves the legislator by extension. But then, should our calculus change if the legislator has been elected for a life term? If so, we commit ourselves to providing a different justification for why the life-term legislator is affected by the system. Maybe the legislator is a stakeholder because they could one day be subjected to the system’s outputs. But then, is everyone in society a stakeholder? And does stakeholderness scale with the degree to which one is likely to be affected by the system? In that case, a corrupt legislator might be considered a more central stakeholder than a law-abiding one, by virtue of being more likely to be affected by the system.

The normative consequence here can be further elucidated by leveraging one of thought experiments’ major strengths, enabling us to go beyond what is \textit{empirically} available to the realm of what is \textit{logically} available. Suppose we elaborate $P$'s role such that $P$ has a relationship to technology $T$ that does not allow any interaction or effect. 

We introduce the \textbf{divine alien legislator} (DAL): a being legitimately elected by the people of Earth to make policy decisions for them. The divine alien legislator lives on a faraway planet several light-years away and, therefore, \edit{its livelihood} cannot be affected by any system on Earth. \edit{We can take this unaffectedness one step further: the DAL is indifferent to what happens on Earth and, therefore, has no personal stake in their own decisions.} However, they can hear and provide their divine perspective on any policy questions, including ones relating to the deployment of technological systems. By many reasonable normative accounts, we might conclude that a legitimately elected divine alien legislator is a stakeholder, too---and that we should, therefore, consider their perspective. And yet, accepting VSD’s conceptualization precludes us from reaching this conclusion. In this case, our commitment is that legitimacy does not matter for stakeholdership. The only attributes that matter are the interaction and effect between technology $T$ and person $P$. Therefore, if we argue that someone is a stakeholder, we have to contend it by virtue of the system's impact on them; we cannot appeal to legitimacy.

The DAL thought experiment enables us to consider a perfect case that does not exist in the real world---that of a legislator who is not (indeed, \text{cannot be}) affected by a system. In doing so, we can bypass arguments about whether legislators are or are not affected by a system to reveal that VSD emphasizes interaction and effect to exclude other potentially influential factors, such as legitimacy. We can then directly ask whether this normative consequence is one we wish to accept.

\section{Testing Alternative Stakeholder Formulations}\label{sec:alt-stakeholder}
The above thought experiments reveal problematic normative consequences of the VSD conceptualization of a stakeholder. Next, we provide two reformulations of a stakeholder and validate these conceptualizations with thought experiments. We continue with the same experimental setup as before, considering a recidivism-risk assessment technology $T$ and experimenting with the social role of person $P.$ However, to better align the normative implication of the conceptualization, we adjust our definition of stakeholder inclusion and the implications of being a stakeholder.

\subsection{Further Bounding Stakeholders}
Reflecting on $P$ as a for-profit prison owner or a divine alien legislator, we need a boundary on stakeholder inclusion that excludes those with solely capital interests but includes those who have legitimate influence over $T$. Fortunately, the stakeholder concept is not original to this paper nor HCI. Therefore, we can turn to previous attempts to bound the concept in Organizational Sciences and use thought experiments to uncover the utility of those boundaries. 

\begin{figure}
    \centering
    \includegraphics[width=\textwidth]{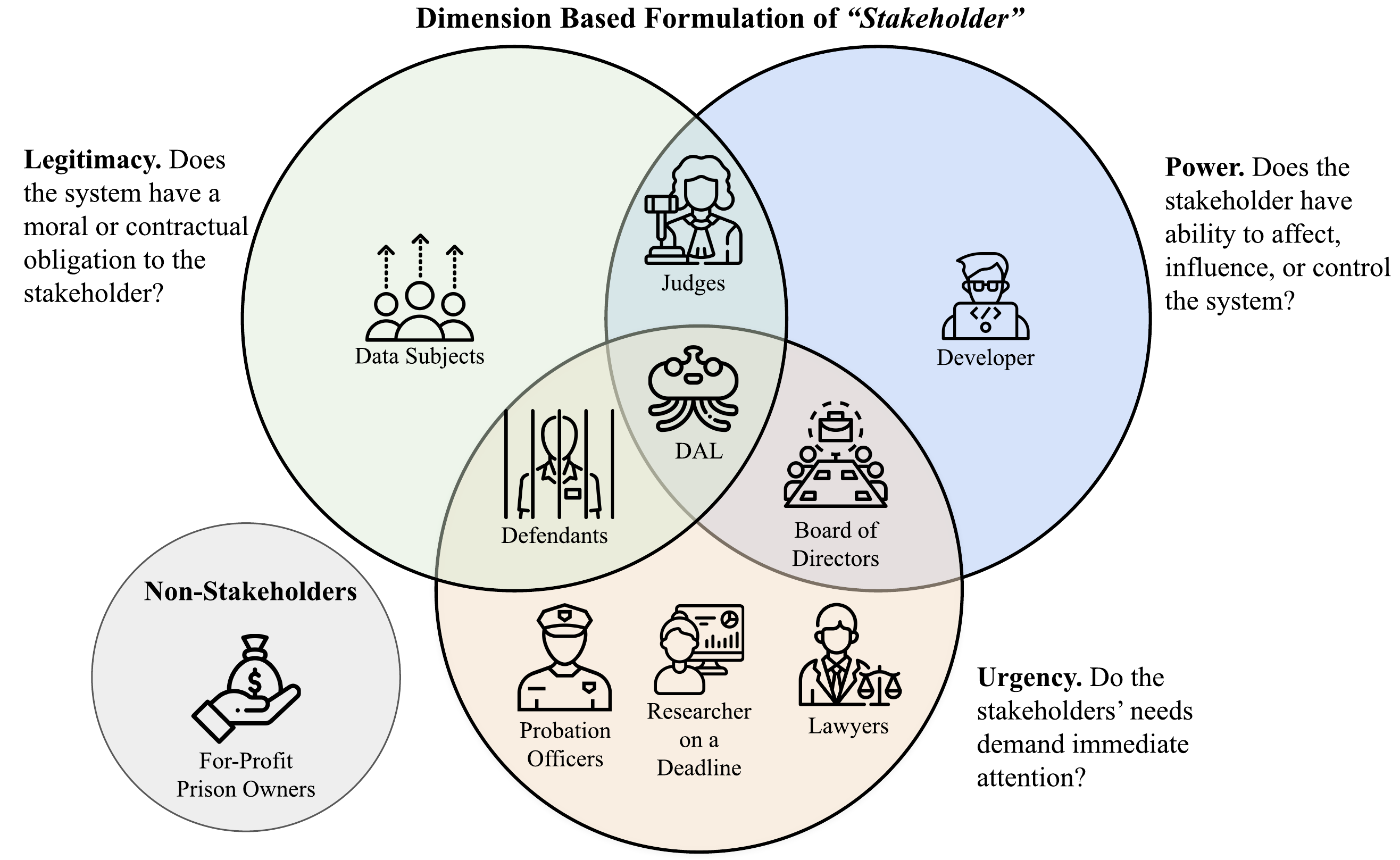}
    \caption{\edit{A dimension-based formulation of stakeholder which captures stakeholder legitimacy, power, and urgency. Figure adapted from~\citep[p. 874]{Mitchell1997-tg}. Note that these dimensions allow us to create a hierarchy of stakeholders. ~\citet{Mitchell1997-tg} call stakeholders who only have urgency (e.g., researchers on a deadline) \textit{``mosquitoes buzzing in the ears"} and note they should not be prioritized over those with legitimacy or power.}}
    \Description{Fig 3. A three-circle venn diagram denoting areas of: (1) legitimate stakeholders (e.g., data subjects), (2) powerful stakeholders (e.g., developers), and (3) urgent stakeholders (e.g., researchers on deadlines). The overlapping spaces depict: defendants as legitimate and urgent stakeholders, judges as legitimate and powerful stakeholders, the board of directors as powerful and urgent stakeholders. The center overlapping point contains the divine alien legislator who is legitimate, powerful, and urgent.}
    \label{fig:dimension-formulation}
\end{figure}
Taken from~\citet{Phillips2003-fn}, stakeholder legitimacy asks: \textit{Does the system have some moral or contractual obligation to the stakeholder?}~\citet{Mitchell1997-tg} expand this notion of stakeholder legitimacy to create a taxonomy of stakeholders that incorporates three salient dimensions: (1) Legitimacy, (2) Power, and (3) Urgency. They establish that being a stakeholder implies one has at least one of these three dimensions. Moreover, these attributes allow us to create a hierarchy of stakeholder considerations when overlaid on top of each other (see Figure~\ref{fig:dimension-formulation}). For example, if $P$ were a researcher on a tight paper deadline studying recidivism predictive systems, they probably have significant urgency but no power or legitimacy.~\citet{Mitchell1997-tg} call stakeholders who only have urgency \textit{``mosquitoes buzzing in the ears"} and note they should not be prioritized over those with legitimacy or power. 

We demonstrate a thought experiment with a legitimacy-centered formulation of stakeholders. In scoping whose perspective we value to legitimate stakeholders only, we can appropriately narrow our conceptualization to the following premises~\edit{for stakeholderism:}

\begin{proof}\leavevmode
    \begin{enumerate}
    \item $P \in \{legitimate\ stakeholders\} \iff (T \text{ has a moral or contractual obligation to } P)$ \hfill(Definition)
    \item $P \text{'s perspective should be valued } \iff P \in \{legitimate\ stakeholders\}$ \hfill (Implication)
    \end{enumerate}
\end{proof}

Now, we can use a thought experiment to test whether this new conceptualization aligns with normative prescriptions. Returning to $P$ as a for-profit prison owner, we have previously established that $P$'s business would be heavily affected by $T.$ However, $T$ has no moral obligation to $P$ beyond the basic moral obligation that $T$ has to all people, such as protecting basic human rights. Moreover, $T$ has no contractual obligation to $P.$ For example, $P$ is not a client paying for $T.$ Therefore, $P$'s perspective should not be valued in the design process. In formal argumentation:

\begin{proof}\leavevmode
    \begin{enumerate}
    \item $P := \text{for-profit prison owner}$\hfill(Given)
    \item $P \in \{legitimate\ stakeholders\} \iff (T \text{ has a moral or contractual obligation to } P)$ \hfill(Definition)
    \item $P \text{'s perspective should be valued } \iff P \in \{legitimate\ stakeholders\}$ \hfill (Implication)
    \item $T \text{ has no obligation to }P$\hfill(Given)
    \item $P \notin \{legitimate\ stakeholders\}$\hfill(From 2,4)
    \item $P$\text{'s perspective should not be valued}\hfill(From 3, 5)
    \end{enumerate}
\end{proof}

The above thought experiment concludes that a designer has no moral imperative to value the perspective of the for-profit prison owner. Practitioners may consider the stances of those who do not have legitimacy, but now we have a vocabulary and logical argument to evaluate the normative consequences of doing so.

Let us consider the other outstanding case: the divine alien legislator. We have previously established that $P$ neither interacts with nor is affected by $T.$ However, $P$ was legitimately elected by the people to create and iterate upon policy, including policy surrounding the use of $T$. In fact, $P$ has been ethically bestowed full decision-making power on whether $T$ can be used for its intended purpose. Given this scenario, $P$ is a legitimate stakeholder. Therefore, $P$'s perspective should be valued.

\begin{proof}\leavevmode
    \begin{enumerate}
    \item $P := \text{divine alien legislator}$\hfill(Given)
    \item $P \in \{legitimate\ stakeholders\} \iff T \text{ has a moral or contractual obligation to } P$ \hfill(Definition)
    \item $P \text{'s perspective should be valued } \iff P \in \{legitimate\ stakeholders\}$ \hfill (Implication)
    \item $P \text{ was legitimately elected by the people for the explicit purpose of creating policy}$\hfill{(Given)}
    \item $P \in \{legitimate\ stakeholders\}$\hfill(From 2, 4)
    \item $P$\text{'s perspective should be valued}\hfill(From 3, 5)
    \end{enumerate}
\end{proof}

By introducing a dimension of legitimacy, we have bounded our conceptualization of stakeholders to solve the under-inclusivity problem revealed when $P$ is a divine alien legislator. The above thought experiments are examples of how to include bounding attributes into a conceptualization and then further test its normative consequences.

\subsection{Adding an Evaluative Normative Theory}
\begin{figure}
    \centering
    \includegraphics[width=\textwidth]{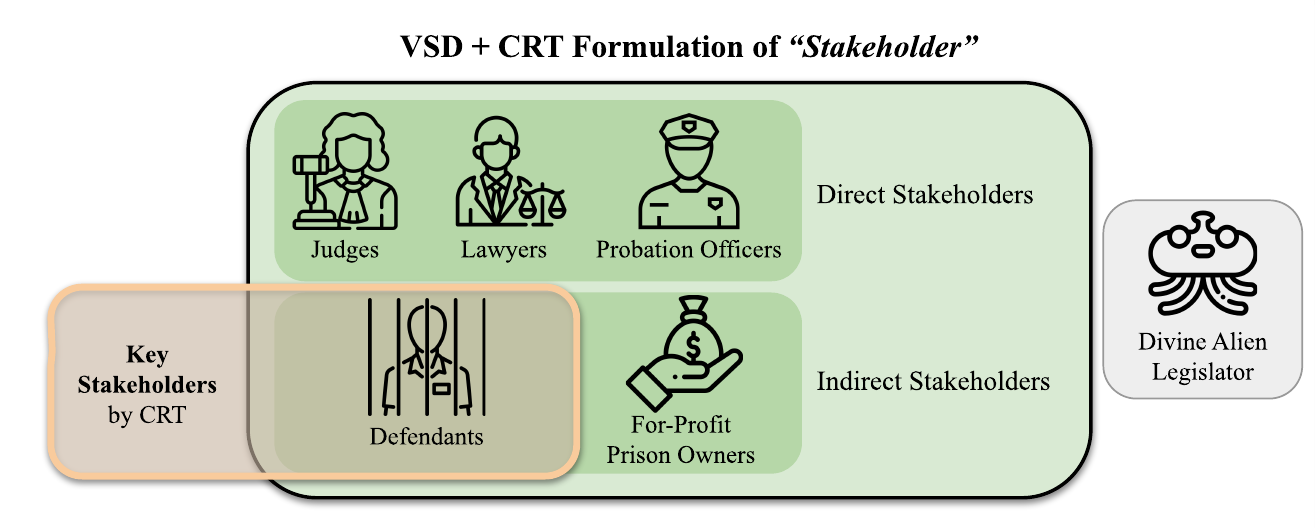}
    \caption{Visual of Value-Sensitive Design's formulation of stakeholders with a Critical Race Theory (CRT) lens. Note that this formulation maintains direct and indirect stakeholders, but prescribes a robust boundary around who can be a key stakeholder.}
    \Description{Fig 4. Visual of VSD's formulation of stakeholders with a CRT lens. Image depicts judges, lawyers, and probation officers as direct stakeholders. Defendants and for-profit prison owners are separated as indirect stakeholders. The divine alien legislator is depicted outside the bucket of stakeholders. Defendants are highlighted with an orange box as key stakeholders by critical race theory.}
    \label{fig:crt-formulation}
\end{figure}

Recall that VSD leaves the onus on the designer to determine key stakeholders---those whose perspectives should be prioritized during the design process. However, HCI has not yet coalesced on robust ways to identify and justify the designer's choices. Therefore, it is imperative we have methods that allow us to normatively judge stakeholder inclusion reasoning. Critical Race Theory (CRT) for HCI~\cite{Ogbonnaya-Ogburu2020-xw} states that researchers are obligated to prioritize the perspectives of systemically marginalized or vulnerable populations (see Figure~\ref{fig:crt-formulation}). By justifying the decision to prioritize marginalized voices in design processes, we can ground the following normative position: empirical methods that elucidate the perspectives of minority groups should be preferred to those that forego or minimize those perspectives. We can incorporate CRT into a thought experiment that adds an evaluative layer to stakeholder inclusion decisions.

Let $P$ be a previously convicted criminal in a minority racial group. Returning to our VSD conceptualization of stakeholder, $P$ is a stakeholder of $T$ if and only if $P$ interacts with $T$ or is affected by $T.$ $P$ is clearly affected by $T$ as $T$'s output directly determines $P$'s sentencing. However, what happens if $P$'s perspective conflicts with the for-profit prison owner who, under this conceptualization, is also a stakeholder? We need a normative framework to justify our decision on stakeholder hierarchy. We can adopt Critical Race Theory, which states that key stakeholders are those who are part of systemically marginalized communities within $T$'s problem space. By this premise, $P$'s perspective should be prioritized over other stakeholders because $P$ is part of a marginalized community. More formally:

\begin{proof}\leavevmode
    \begin{enumerate}
    \item $P := \text{previously convicted criminal in a minority racial group}$\hfill(Given)
    \item $P \in \{stakeholders\} \iff (P \text{ interacts with } T \lor P \text{ is affected by } T)$ \hfill(VSD-definition of stakeholder)
    \item $P \text{'s perspective should be valued } \iff P \in \{stakeholders\}$ \hfill (VSD-implication of stakeholder inclusion)
    \item $ P\text{'s perspective should be prioritized if } P \text{ is a systemically marginalized stakeholder} $\hfill(By CRT)
    \item $P \text{ is affected by } T$ \hfill(Given)
    \item $P \in \{stakeholders\}$ \hfill(From 2, 5)
    \item $P$\text{'s perspective should be valued}\hfill(From 3, 6)
    \item $P$\text{'s perspective should be prioritized}\hfill(From 1, 4) 
    \end{enumerate}
\end{proof}

Through this thought experiment, we have justified the decision to prioritize marginalized voices in design processes. Furthermore, this justification explicates the normative framework--Critical Race Theory--that we are adopting. Readers can explicitly understand our philosophical stance rather than tacitly reasoning about it. Incorporating CRT is normative insofar as it describes a moral obligation; when we elaborate on the concept of a key stakeholder with Critical Race Theory, we establish a moral hierarchy of perspectives (see Figure~\ref{fig:crt-formulation}). This allows designers to justify design decisions that they made for stakeholder inclusion.

\subsection{Evaluating our Thought Experiment}
Next, we evaluate these thought experiments based on our criteria outlined in Section~\ref{sec:how-to}.

\subsubsection{Analogy.}
All of our thought experiments rely on the same setup: a recidivism-risk predictive system, $T$, to be used by judges in courts. This setup parallels emerging technologies. In 2016,~\citet{Angwin2016-jy} audited COMPAS, a recidivism-risk system used in Wisconsin courts. They found the system disproportionately predicted future criminal activity along racial lines. 

Furthermore, the experimental variables we chose have parallels to stakeholders that designers must consider. In our case, judges represent the group of intended users; the for-profit prison owner represents stakeholders who primarily have capital interests in the system. While the divine alien legislator is an absurd hypothetical, legislators are often asked to dictate policies that they will never be directly affected by. For example, male legislators can vote on current legislation around women's health. We chose to use the divine alien legislator rather than a real-world example because it focuses the reader's attention on a stakeholder's level of interaction and effect. While our thought experiment may not be realizable, it is tractable.
\subsubsection{Argument.} Moreover, thought experiments should be assessed in terms of logical validity and soundness.

\textit{Validity. }We present our thought experiments in prose and formal logic to highlight their logical validity. We rely on the logical implications from the initial premises to reach our conclusions. For example, VSD states that $P$ is a stakeholder if and only if they interact with technology $T$ or are affected by $T$. We use this if and only if to determine that judges are stakeholders in our experiment and, therefore, judges' perspectives should be valued.

\textit{Soundness. }Our logical representation of VSD stakeholders is heavily informed by foundational VSD literature~\cite{Friedman1996-kx}. Therefore, we argue that our formulation is sound. Note that VSD relies heavily on practitioners identifying key stakeholders (See Figure~\ref{fig:vsd-formulation}). VSD stakeholder theory demands breadth for this practitioner's discretion. However, we show that this conceptual breadth leads to normative breadth; VSD does not ground normative standards on who should be prioritized in design. Even though we take a broad representation of VSD's position on stakeholder theory, we believe it accurately represents the level of practitioner discretion within VSD methodologies.

\section{Generating Formulations and Concepts}\label{sec:generating-formulations}
The previous sections have demonstrated an interrogative application of thought experiments—we take a widely accepted conceptualization of stakeholders, and then interrogate and iterate upon it. In this section, we ask:
\begin{quote}
    \textit{Can we use thought experiments to speculate about a concept that does not have a clear formulation, but is the center of emerging discussions?}
\end{quote}
Recall that a conceptualization serves as an initial proposition to interrogate. Therefore, thought experiments may also be generative in defining and clarifying new concepts in HCI and related areas. In this section, we present thought experiments where the goal is to generate a concept or conceptualization rather than interrogate one.

We highlight Helen Nissenbaum's~\cite{Nissenbaum2004-og, Nissenbaum2011-fz, Barth2006-bu} use of thought experiments as a conceptually generative method for HCI. Specifically, we formalize the thought experiments used by Nissenbaum and her peers to create the building blocks of contextual integrity. Contextual integrity asserts that privacy is about situational information norms rather than simply keeping certain information secret. Contextual integrity has been a staple framework in the HCI community for considering moral obligations surrounding personal data and information~\cite{Chancellor2019-cz, zimmer2018addressing, Skeba2020-zv}.

While Nissenbaum often used thought experiments as an explanatory device, our thought experiment structure from Section~\ref{sec:how-to} elucidates the formulations produced through this conceptual work. There is an ongoing discourse about applying contextual integrity to emerging machine learning methods in HCI and related human-centered areas~\cite{Chancellor2019-cz}. Below, we apply our thought experiment structure to three thought experiments used by Nissenbaum and her peers. We illustrate how these thought experiments were generative as they created appropriate conceptual boundaries for the normative arguments Nissenbaum sought to make. Notably, these thought experiments were successful as contextual integrity has since informed radical changes in privacy policy~\cite{Madrigal2012-is}. 

\subsection{Basic Example}
\begin{quote}
    \textit{One further feature is key to understanding what we mean here by “contexts,” for not only are they characterized by roles and norms but also by certain ends, or values. In the case of health care, an onlooker (say, from another planet) observing a typical health care setting of a hospital, will be unable to make proper sense of the goings-on without appreciating the underlying purpose behind it, that is, alleviating illness and promoting health. Although settling the exact nature of the ends and values for any given context is not a simple matter—even in the case of health care, which is relatively robust—the central point is that the roles and norms of a context make sense, largely, in relation to them.}~\citet[p. 3]{Barth2006-bu}
\end{quote}

Crucial to the notion of contextual integrity, is the concept of a ``context." However,~\citet{Nissenbaum2004-og} notes that context is not well-defined in most disciplines, including HCI~\cite{dourish2004we}. Therefore, we must generate a formulation of context before establishing contextual integrity. We take the thought experiment from ~\citet{Barth2006-bu} and demonstrate how it proves generative. 

\textbf{Analogy. }Imagine we have hospital $H$ as a representation of a typical healthcare setting. Like most hospitals, $H$'s core goals are to alleviate illness and promote health. Now, suppose person $P$ is observing $H$.

\textbf{Experimental Variable. }Let the experimental variable be $P$'s understanding of $H$'s goals. Therefore, we have the following cases:
\begin{enumerate}
    \item Let $P$ be a visiting doctor. Therefore, $P$ has a deep and internalized understanding of $H$'s goals. In fact, $P$ has agreed to the Hippocratic Oath, and, therefore, has deliberately aligned their values to the values of $H$. In this case, we can conclude that $P$ fully appreciates the purpose of $H$.

    \item Let $P$ be an average person who stops and observes the hospital through a window. While $P$ is seeing everything a doctor may see, $P$ may not have the same goals as a doctor. However, $P$ has been educated on the goals of hospitals and the risks of a world without health care. $P$ has probably been sick before and needed to go to the hospital themselves. In this case, $P$ may not professionally align with $H$'s goals but certainly appreciates them.

    \item Let $P$ be an alien from a planet with no sickness and, therefore, no hospitals. $P$ is given a looking glass where they can observe all of the ongoings of $H$. In this case, $P$'s understanding of the situation is distinctly different from a human understanding because they do not have an appreciation for the underlying purpose of a hospital.
\end{enumerate}

\textbf{Conclusion. }This thought experiment suggests that context requires a relational conceptualization. Rather than an objective set of roles and norms (i.e., the jobs and tasks of a hospital), there is also a level of values-understanding that contributes to context. In this sense, context is a set of roles and norms that make sense given the values of the settings.~\citet{Barth2006-bu} use this idea that contexts are distinctive social settings with underlying ends to formalize the notion that contexts contain norms. The violation of these contextual norms contributes to a violation of privacy under contextual integrity.

\subsection{Generating a Formulation Based on A Counterfactual}
\begin{quote}
\textit{A second consideration is the compatibility of notice-and-consent with the paradigm of a competitive free market, which allows sellers and buyers to trade goods at prices the market determines. Ideally, buyers have access to the information necessary to make free and rational purchasing decisions. Because personal information may be conceived as part of the price of online exchange, all is deemed well if buyers are informed of a seller’s practices collecting and using personal information and are allowed freely to decide if the price is right...A  deeper ethical question is whether individuals indeed freely choose to transact–accept an offer, visit a website, make a purchase, participate in a  social network–given how these choices are framed as well as what the costs are for choosing not to do so. While it may seem that individuals freely choose to pay the informational price, the price of not engaging socially, commercially, and financially may in fact be exacting enough to call into question how freely these choices are made. }~\citet[p. 34]{Nissenbaum2011-fz}
\end{quote}
In this thought experiment, we start with the following conceptualization:

\textbf{Proposition. }All exchanges of personal information for online resources (i.e., information flows) are an instantiation of a free market exchange.

\textbf{Analogy. }Suppose we have person $P$ with personal information $P_{i}$ using online resource $O$ in a free market setting. In essence, $P$ is a buyer looking to exchange $P_{i}$ for use of $O$.

\textbf{Experimental Variable. }Let the experimental variable be the consequence to $P$ of rejecting the exchange of $P_{i}$ for $O$. Therefore, we have the following cases:
\begin{enumerate}
    \item Let $P_{i}$ be $P$'s credit card information and $O$ be an online store. In this case, if $P$ rejects the exchange, $P$ cannot purchase an item from $O$. This consequence seems reasonable as it mirrors an off-line setting; in order to buy most things an individual has to give their credit card information to the storekeeper.
    \item Let $P_{i}$ be $P$'s cookie activity and $O$ be an informational website on Medicare, a government-funded healthcare plan. In this case, if $P$ rejects the exchange, $P$ risks not getting necessary public health information.
    \item Let $P_{i}$ be all of $P$'s activity on social media platform $O$. Moreover, $P$ uses $O$ specifically to connect with recovery support groups. Now the consequence is extreme; if $P$ rejects the exchange then they lose access to a necessary support group for their livelihood. 
\end{enumerate}

\textbf{Conclusion. }From these cases, we can conclude that there exists an information flow that is not a free exchange. In Case 2, $P$ no longer has access to a public information resource. In Case 3, $P$ no longer has access to a necessary support group. This thought experiment suggests that the privacy paradigm captured by notice-and-consent is based on a faulty presumption that information exchanges mirror free-market exchanges of goods. Now we can ask: Which dimensions would a new conceptualization of privacy need to include? Our experimental variable was the consequence of $P$ rejecting an information exchange. We captured various cases by changing the values of $P_{i}$, $O$, and $P$'s intended use of $O$. This suggests that a new conceptualization of privacy would need to be bounded by the following dimensions: information type, recipient type, and use norms. In fact, Nissenabum incorporates these three dimensions as a subset of the five parameters of contextual integrity: (1) the information subject; (2) the sender; (3) the recipient; (4) the information type; and (5) the transmission principles which represent the norms of information flow~\citep[p. 3]{Benthall2017-rw}. A contextual integrity analysis requires a practitioner to define each of these parameters in relation to the context they are assessing.

\subsection{Generating a Concept Based on an Emerging Event}
\begin{quote}
\textit{Remember the hubbub over Google Street View in Europe? Germans, in particular, objected to the photo-taking cars. Many people, using the standard privacy paradigm, were like, "What's the problem? You're standing out in the street? It's public!" But Nissenbaum argues that the reason some people were upset is that reciprocity was a key part of the informational arrangement. If I'm out in the street, I can see who can see me, and know what's happening. If Google's car buzzes by, I haven't agreed to that encounter. Ergo, privacy violation.}~\citet[p. 2]{Madrigal2012-is}
\end{quote}

In the early 2010s, Google deployed numerous labeled cars with cameras to take photos for Google Street View, a new feature of Google Maps\footnote{https://www.google.com/streetview/how-it-works/}. Subsequently, a German woman sued Google for taking pictures of her house, arguing that Googled violated her privacy~\cite{McCarthy2011-lf}. However, the court ruled that taking photos from the street is legal, as streets are public property rather than private. While this ruling was based on the dichotomy of public vs. private, we ask: What concept is being violated here that is not violated in similar situations?

\textbf{Analogy. }Suppose $P$ is an average German citizen on a German street.

\textbf{Experimental Variable. }Our experimental variable here is visibility. We present the following cases:
\begin{enumerate}
    \item Suppose we are in a world without cameras and $P$ is out taking a walk. In this case, $P$ is only visible to the other people on the street. Notably, if someone on the street can see $P$, $P$ can likely see them too.

    \item Suppose $P$ is on a walk and ends up in the background of a tourist photo taken by tourist $T$. $T$ may show the photo to their friends and maybe even post it on social media. In this case, $P$ cannot see the audience of $T$'s friends that will end up seeing the photo.

    \item Suppose $P$ walks in front of the Google Street View car and, therefore, is in a photo on Google Maps. Similar to Case 2, $P$ cannot see the audience that can see $P$. However, this metaphorical audience is now all Google Maps users who use the Google Street View feature on this German street. Notably, this is potentially orders of magnitude larger than the audience in Case 2.
\end{enumerate}

\textbf{Conclusion. }This thought experiment demonstrates that the incorporation of cameras into society changes an individual's visibility. Moreover, the addition of the Google Street Car adds a dimension of scale to the scenario. This combination of visibility and scale is what Nissenbaum uses to formulate the concept of reciprocity: ``\textit{is it possible for subjects to see those who see them}"~\citep[p. 229]{Nissenbaum2019-cr}. While reciprocity does not need to be equivalent, technology causes it to fracture significantly. Nissenbaum incorporates this concept of reciprocity into a larger privacy framework by thinking of the (bi)-directional norms of information flows~\cite{Barth2006-bu}. For example, there are no norms around information reciprocity between a patient and their doctor; doctors should not share their personal medical state with their patients. However, friends typically exchange phone numbers in a bidirectional manner. Reciprocity allows us to view information flows as exchanges with direction and scale, rather than a simple transfer from one person to the other. Notably, even though the German courts ruled in Google's favor, Google placed a voluntary 10-year halt on its Street View program in Germany after German public outcry at the court's decision~\cite{Cantrill2023-hz}.

The above examples highlight how thought experiments can be conceptually generative in addition to being interrogative. In each thought experiment, the conclusions lead us to a new formulation that is essential to building out contextual integrity. For example, understanding that visibility also has a component of scale (the size of the resulting audience) allows us to generate the concept of \textit{reciprocity}. Future work in HCI can use this generative capacity of thought experiments to build out new concepts surrounding emerging topics. For example, tangential to data privacy, recent work has focused on theorizing data as labor to roadmap future research directions but still lays out a myriad of open questions~\cite{Li2023-wa}. We suggest that thought experiments can precisely build out foundational concepts surrounding data contribution, labor, and leverage in the same way Nissenbaum does for data privacy.

\section{Discussion}\label{sec:discussion}
We introduce thought experiments as a method for interrogating the normative consequences of conceptual commitments that HCI researchers make. This section discusses the utility of thought experiments to~\edit{expand HCI's methodological toolbox}. We discuss concepts beyond stakeholders that could benefit from further application of our method.~\edit{Finally, we} identify key opportunities for researchers, reviewers, and critics to apply thought experiments.

\subsection{Progressing the Methodology of HCI}
Conventional methods and theories in HCI are well suited to answering certain questions: \textit{Which of these two design variants will balance best among the desired criteria~\cite{Norman2013-rr}?} \textit{What are the processes by which users come to adopt (or reject) a certain software~\cite{Grudin1988-me}?} However,~\edit{as technology becomes part of our everyday lives}, recent HCI work has begun engaging with moral questions: \textit{What would it look like to adopt normative lenses, such as feminist theory~\cite{Bardzell-2010-FeministHCITaking}, Critical Race Theory~\cite{Ogbonnaya-Ogburu2020-xw}, or FAccT\footnote{Fairness, Accountability, and Transparency} frameworks~\cite{Lepri2018-rt}, to achieve a moral standard?}

~\edit{While these theories in HCI allow us to justify moral stances, HCI desperately needs methods that interrogate how practitioners take on implicit normative stances.} For instance, how we conceptualize ``the user" has consequences in focusing our research, design, and policy attention on certain actors and unintentionally ignoring others~\cite{Baumer2017-gc}. These conceptualizations also have an impact on who matters, not just for the field of HCI but for the much broader space of compassionate design in sociotechnical systems. The above thought experiments illustrate an analogous point for the concept of stakeholders: that different formulations of the concept carry different implicit commitments with different normative consequences.

We take the stance that understanding~\edit{these unstated normative consequences} is crucial to understanding what makes harm possible in sociotechnical systems. For example, what makes it possible for certain stakeholder groups, such as data subjects, to be consistently under-considered in predictive systems~\cite{Li2023-wa}?~\edit{Our thought experiments demonstrate that VSD presumes a broad definition of stakeholderism. This methodological limitation has moral implications for whose voice gets considered in the design process. By recentering stakeholderism around morally-laden criteria, such as legitimacy or marginalization, we can make these implicit normative positions explicit.} 

Our work is in conversation with previous scholarship surrounding marginalization and harms. For example,~\citet{Angwin2016-jy} ask: \textit{What makes it possible for predictive systems to propagate human bias?}~\citet{Fricker2007-zh} asks: \textit{What makes it possible for systems of research to propagate the silencing and exclusion of certain voices?} While thought experiments are not a silver bullet to answering these questions, they illuminate where conceptual commitments are laden with values. Our thought experiments show that uninterrogated conceptual commitments can make the morally suspect seem benign, despite researchers' intentions.

Bringing these normative implications to light is essential to progressing HCI towards~\edit{its ethical aspirations}. For example,~\citet{Chancellor2019-cz} found that by articulating ethical tensions in predictive mental health, they revealed the methodological gaps that amplified the normative problems plaguing the field.~\citet{Bruckman2017-qq} posit that community-based norms are the foundation for research ethics. We use thought experiments to unpack how certain formulations can contribute to~\edit{community-wide} values misalignments, such as misidentifying key stakeholders. We show that further thought experiments can reformulate concepts to reduce misalignment. In other words, thought experiments answer: \textit{What makes it possible for systems to be so misaligned?} Interrogating these questions allows us to validate potential conceptual solutions.

\edit{Our proposal to progress the methodology of HCI presumes that practitioners are equipped to use thought experiments. We believe this presumption is justified, as there is a rich history of domain experts (i.e., physicists, computationalists, and mathematicians) engaging in thought experiments without strict philosophical training. However, there may be a learning curve to the widespread adoption of thought experiments in HCI and related research venues (CHI, CSCW, FAccT, etc.). Future efforts could leverage pre-existing infrastructure, such as educational workshops at conferences and symposia, to engage with external experts. The authors of this paper are committed to fostering those efforts where necessary, as we believe in sustained and interdisciplinary efforts to increase the quality of ethical reasoning in research.}

\subsection{Future Subjects for Thought Experiments}
We use the concept of stakeholders in HCI as an instance of conceptual work that has led to uninterrogated normative consequences. This section highlights other concepts that could benefit from thought experiments. Future work could apply our thought experiments to analogous problem areas or leverage the method outlined in this paper to innovate new thought experiments.

Stakeholder inclusion has become a specific problem in~\edit{Artificial Intelligence (AI)} pipelines, leading to ``participatory AI" methods~\cite{Delgado2021-ec}. Similar to VSD, these methods make conceptual commitments about who is a participant and the nature of participating. Properly conceptualizing AI participation is crucial in the current zeitgeist. Many communities have taken drastic action as they have been excluded from participating in AI, despite their data contributions. For example, fanfiction communities are poisoning generative AI models~\cite{Silberling2023-yx}, and screenwriters are on strike because of job threats from ChatGPT~\cite{Wilkinson2023-mr}. Our work on interrogating conceptualizations of inclusion aligns with current calls for reconceptualizing data contributors as laborers~\cite{Li2023-wa} and AI advocates as participants~\cite{Queerinai2023-xy}. Future work could adapt our thought experiments to the rapidly emerging field of generative AI. For example, if we changed our technology $T$ to a~\edit{generative AI system}, what are the normative consequences of appealing to stakeholder legitimacy?

Moving away from questions about participant inclusion, thought experiments may be useful when generating idealistic concepts. For example, Helen Nissenbaum's work highlights the methods gap that thought experiments solve; her work was not originally published in HCI venues, despite being heavily cited in HCI papers today~\cite{zimmer2018addressing, Skeba2020-zv, Chancellor2019-cz}. Speculatively, this is because Nissenbaum used non-traditional methods for HCI at the time (i.e., thought experiments). When we have a subjective concept, such as privacy, how can we reach a formal and robust conceptualization? Nissenbaum's work frequently uses thought experiments to demonstrate how viewing privacy as contextual integrity demands that information flows cannot be evaluated in a vacuum. We argue that HCI does not have a formal way of evaluating these hypothetical but tractable logical arguments. This gap can potentially lead to further exclusion of necessary work, such as Nissenbaum's. Future work could continue Nissenbaum's conceptual inquiry into privacy. For example, researchers could build on~\citet{Skeba2020-zv} and ask: \textit{What are the normative consequences of committing to contextual integrity in inferential systems?} By answering these questions, HCI practitioners can precisely link value-laden conceptual decisions to their implications, allowing us to build and design more ethical systems.

\subsection{Thought Experiments as a Tool For...}
\edit{Thought experiments originated from philosophy. However, they have a rich history of being used by scientists to answer the philosophical questions of their field. For example, Newton and Galileo were physicists; Pythagoras was a mathematician; Alan Turing was a computer scientist. While some discussions about thought experiments are technical~\cite{Gendler1998-es, Norton2002-av}, thought experiments have never been a technical method reserved for philosophy experts. In fact, HCI researchers are uniquely qualified to engage with thought experiments, given our expertise lies at the intersection of theoretical and technical domains. We propose thought experiments to build on HCI researcher expertise while addressing a methodological gap in the field. his section outlines key opportunities for HCI researchers to engage with thought experiments as tools for supplementing their other work.}

\subsubsection{Innovators in HCI}
Innovation in HCI is often motivated by either (1) anticipating user needs or (2) reflecting on failures to meet user needs.~\citet{Ackerman2000-ik} articulated the theory of sociotechnical gaps by identifying failed applications. Thought experiments can similarly identify moments for creative problem-solving and beneficial reconceptualizations. In this paper, we demonstrate how HCI can use thought experiments to build appropriate guardrails around design motivations. This mirrors the success of frameworks and taxonomies, which often serve as comprehensive lists of the status quo to guide future design~\cite{Oulasvirta2016-en}. Thought experiments iterate on the concepts that underlie these tools. For example, we show asking ``what are the consequences of stakeholderness?'' raises the question ``who is a stakeholder?'' Moreover, we show how to identify concepts that need to be strengthened. Recall our thought experiment about including a for-profit prison owner as a stakeholder. Here, we appeal to the hypothetical nature of thought experiments to identify conceptual weakness. However, it raises a larger gap within innovative HCI---how can we innovate with weak concepts? Innovation in HCI relies on appropriately scoping how we motivate design resources. Flawed conceptualizations risk misappropriating our attention to the point where we may miss the mark entirely. Therefore, we propose thought experiments as a way of scoping design motivation and, subsequently, foster more focused innovation in HCI.


\subsubsection{Critics in HCI}
Second, thought experiments can ground criticism around emerging technologies, not only in academia~\cite{DeVito2021-yv, Cheng2022-dv} but also in the numerous reports of harms caused by unchecked technologies. With thought experiments, we can focus our interrogation on specific dimensions, such as stakeholder power, and bring normative consequences to the forefront of the conversation. For example, our divine alien legislator demonstrates the importance of considering stakeholder legitimacy. Without considering legitimacy, we risk prioritizing capital interests, such as the for-profit prison owner, in our design processes. Especially with the advent of AI technology, there is a growing need to rigorously critique large platform technologies. Moreover, many critical theorists argue that we should examine how technology practices intersect with identity, class, and power~\cite{Brock2018-qu, Pierce2015-ww, Ogbonnaya-Ogburu2020-xw}. Thought experiments serve as a complement to the important and ongoing work in criticism. We hope that future critiques use thought experiments to iterate on other prominent concepts in social technology.
 
\subsubsection{Reviewers in HCI}
Thought experiments could also be applied in the peer review process.~\citet{Hecht2021-qx} recently called for normative implications to be an evaluative criterion during peer review. They argue that authors have an obligation to evaluate the potential negative consequences of their work; reviewers, in turn, have an obligation to scrutinize said consequences when considering the merits of the work. Thought experiments can be incorporated into the review process as a mechanism for authors and reviewers to explore, defend, or critique broader implications of potential publications. For example, authors who study non-traditional communities may use thought experiments to justify broader stakeholder inclusion. Reviewers may use a thought experiment to critique an author's assumptions around stakeholderness and moral hierarchy, thereby giving a reviewer grounds to reject a paper on flawed normative reasoning. Thought experiments open the door to a transparent and precise dialogue surrounding the positive and negative normative consequences of research projects.

Our work establishes a method to reason about potential broader impacts. Thought experiments tie~\edit{moral intuition} to logical arguments. HCI researchers are often asked to consider their~\edit{empirical findings} alongside their~\edit{informed moral speculations} without clear scaffolding. Venues such as NeurIPS, EMNLP, and the NSF have recently encouraged publications to have broader impact statements. However, prior work has found a large information disparity in these statements~\cite{Ajmani2023-bg, Nanayakkara2021-nm}. Thought experiments can precisely explicate the potential consequences of implicit conceptual commitments. For example,~\edit{Machine Learning (ML)} practitioners could use thought experiments to formally address how they conceptualize data subjects in their pipelines. This application could help address known ethical tensions, such as the privacy implications of inferring mental health states from social media data~\cite{Chancellor2019-cz}; compensation for data subjects who are doing labor~\cite{Li2023-wa}; and copyright implications for non-traditional data sources, such as online fanfiction~\cite{Fiesler2016-fi}.

Thought experiments invite rules into the conversation to focus the reader toward specific dimensions. This makes thought experiments interrogative rather than speculative -- the experiment takes on a stance (conceptual commitment) and seeks to precisely understand its normative implications. Consequently, using thought experiments creates a formal argument about potential harm. Thought experiments are a precise method that relies heavily on logical steps.


\subsection{Limitations}
This paper proposes thought experiments as a conceptual tool for HCI researchers. While this use of thought experiments aligns with previous philosophical work on the method~\cite{Kornberger2020-dv}, it is only one potential use of thought experiments for deliberate thinking. As mentioned, thought experiments have been used as explanatory devices for moral positions (see The Violinist Argument~\cite{Thomson2004-hk}) or quasi-experiments for the natural sciences (see Schr\"{o}dinger's Cat~\cite{Schrodinger1935-tt}). Future work could expand our initial proposal of thought experiments and adapt them to solve non-conceptual problems in HCI.

We also adopt Norton's view that thought experiments are a unique type of argument~\cite{Norton1991-zt}. Although this view is well-accepted, others have argued it is too conservative. For example,~\citet{Gendler1998-es} directly responds to Norton saying that thought experiments generate new knowledge rather than simply revealing certain truths. This debate is ongoing in philosophy; future work could adopt Gendler's (or others') stance to reveal new and exciting uses for thought experiments in HCI.

\section{Conclusion}
This paper proposes thought experiments as a method for conceptual work. We leverage the rich philosophical history of thought experiments to highlight how they can logically take conceptual stances to their normative consequences. \edit{We demonstrate the use of thought experiments as an interrogative tool for borrowed concepts. We validate thought experiments' value to HCI by proposing various thought experiments around conceptualizations of stakeholders.} Thought experiments raise the potential for HCI researchers to better reflect on ethics, broader impacts, and their own positionalities. We suggest that if thought experiments were heavily used in HCI, the field would be better equipped to meet current moral imperatives.

\bibliographystyle{ACM-Reference-Format}
\bibliography{main}

\end{document}